\documentclass[reviewcopy]{elsart}

\usepackage{graphicx}
\usepackage{enumerate}
\usepackage{graphicx}
\usepackage{amsmath}
\usepackage{amsfonts}
\usepackage{amssymb}
\newcommand{\ket}[1]{\left| #1 \right\rangle}
\newcommand{\bra}[1]{\left\langle #1 \right|}

\begin{document}
\begin{frontmatter}
\title{Realization of a holonomic quantum computer in a chain of three-level systems}
\author[x]{Zeynep Nilhan G\"urkan},
\ead{nilhan.gurkan@gediz.edu.tr }
\author[y]{Erik Sj\"oqvist\corauthref{fi}}  
\corauth[fi]{Corresponding author.}
\ead{erik.sjoqvist@kemi.uu.se}
\address[x]{Department of Industrial Engineering, Gediz University, Seyrek, 35665, 
Menemen-Izmir, Turkey \\
Centre for Quantum Technologies, National University of Singapore,
3 Science Drive 2, 117543 Singapore, Singapore} 
\address[y]{Department of Physics and Astronomy, Uppsala University, Box 516, 
SE-751 20 Uppsala, Sweden \\ 
Department of Quantum Chemistry, Uppsala University, Box 518, Se-751 20 Uppsala, Sweden}
\date{\today}
\begin{abstract}
Holonomic quantum computation is the idea to use non-Abelian geometric phases to implement 
universal quantum gates that are robust to fluctuations in control parameters. Here, we propose 
a compact design for a holonomic quantum computer based on coupled three-level systems. 
The scheme does not require adiabatic evolution and can be implemented in arrays of atoms 
or ions trapped in tailored standing wave potentials. 
\end{abstract} 
\begin{keyword}
Geometric phase; Quantum gates  
\PACS 03.65.Vf, 03.67.Lx    
\end{keyword}
\end{frontmatter} 
\maketitle
\section{Introduction}
Holonomic quantum computation (HQC), first proposed by Zanardi and Rasetti \cite{zanardi99}, 
is the idea to use non-Abelian geometric phases to implement quantum gates. In the case of 
adiabatic evolution, this approach allows for universal quantum computation by composing 
holonomic gates associated with a generic pair of loops in the space of slow control parameters. 
Adiabatic holonomic gates are insensitive to random fluctuations in the parameters and therefore 
potentially useful for robust quantum computation \cite{pachos01}. Physical realizations of 
adiabatic HQC have been developed in quantum optics \cite{pachos00}, trapped ions 
\cite{duan01,pachos02} or atoms \cite{recati02}, quantum dots \cite{solinas03,bernevig05}, 
superconducting qubits \cite{faoro03,brosco08,pirkkalainen10}, and spin chain systems 
\cite{wu05,renes13,chancellor13}.

Universal HQC has been demonstrated \cite{sjoqvist12} by using non-adiabatic non-Abelian 
geometric phases \cite{anandan88}. Conceptually, a non-adiabatic holonomy depends on a
loop traced out by a subspace of the full Hilbert space, rather than a loop in some control 
parameter space. An explicit scheme for non-adiabatic HQC, encoding qubits in the two 
bare ground state levels of $\Lambda$-type systems, has been developed in Ref.~\cite{sjoqvist12}. 
This scheme was subsequently realized for a transmon qubit \cite{abdumalikov13}, in a 
nuclear magnetic resonance setup \cite{feng13}, and in a nitrogen-vacany color center in 
diamond \cite{arroyo-camejo14,zu14}.  Furthermore, the idea of non-adiabatic HQC 
has been combined with other methods to achieve resilience to collective errors 
\cite{xu12,zhang14a,liang14,xu14a,xu14b} and has been demonstrated for other level 
structures \cite{mousolou14,zhang14b}. 

Here, we demonstrate non-adiabatic universal HQC in a linear chain of interacting three-level 
systems. The resources scale linearly with the number of logical qubits and can therefore be used 
to build a compact holonomic quantum computer with a small overhead of auxiliary systems. 
Our setup can in principle be implemented for three-level atoms or ions trapped in standing  
wave potentials. 

The outline of the paper is as follows. In the next section, the general idea of non-adiabatic
holonomic quantum computation is described. The model system is introduced in Sec.
\ref{sec:xymodel}. We demonstrate a universal set of one- and two-qubit holonomic gates 
in Sec. \ref{sec:hqc}. While the one-qubit gates in this set are identical to those developed 
in Ref.~\cite{sjoqvist12}, the two-qubit gates differ as they are mediated by ancillary systems 
sandwiched between the logical qubits, rather than by utilizing direct coupling of 
$\Lambda$ systems. The paper ends with the conclusions.

\section{Non-adiabatic holonomic quantum computation}
\label{sec:nahqc}
Let a computational system be encoded in a subspace $\mathcal{S}$ of some Hilbert space 
$\mathcal{H}$. A cyclic evolution of $\mathcal{S}$ implements a quantum gate. This 
gate generally contains a dynamical and a geometric contribution that combine into a unitary 
transformation acting on $\mathcal{S}$. The dynamical part is essentially given by the Hamiltonian 
$H(t)$ projected onto the evolving computational subspace. The geometric part probe the 
underlying geometry of the space of subspaces; technically this space is a Grassmannian 
manifold $G(\dim \mathcal{H};\dim \mathcal{S})\equiv G(N;K)$ \cite{bengtsson06}. 

Non-adiabatic HQC on $\mathcal{S}$ is realized when $P(t) H(t) P(t) = \epsilon (t) P(t)$, 
where $\epsilon (t)$ is the average energy of the subspace at time $t$ and $i\dot{P}(t) = 
[H(t),P(t)]$ with $P(0)$ the projection operator onto $\mathcal{S}$ (we put $\hbar = 1$ 
from now on). For a cyclic evolution, i.e., $P(\tau) = P(0)$, the time evolution operator 
projected onto $\mathcal{S}$ becomes unitary, and we find  
\begin{eqnarray} 
P(0) U(\tau,0) P(0) & = & e^{-i\int_0^{\tau} \epsilon (t) dt} \sum_{a,b=1}^K 
\left( {\bf P} e^{i\oint_C \mathcal{A}} \right)_{ab} 
\ket{\zeta_a (0)} \bra{\zeta_b (0)} , 
\end{eqnarray}
where $\mathcal{A}_{ab} = i\bra{\zeta_a (t)} d \zeta_b (t) \rangle$ is the matrix-valued 
connection one-form with $\{ \ket{\zeta_a (t)} \}$ any orthonormal basis along the loop 
$C$ in $G(N;K)$ such that $\ket{\zeta_a (\tau)} = \ket{\zeta_a (0)}$. Here, 
\begin{eqnarray}
U(C) \equiv \sum_{a,b=1}^K \left( {\bf P} e^{i\oint_C \mathcal{A}} \right)_{ab} 
\ket{\zeta_a (0)} \bra{\zeta_b (0)} 
\end{eqnarray}
is the holonomic gate associated with $C$. The dynamical phase reduces to 
an unimportant overall U(1) phase factor $e^{-i\int_0^{\tau} \epsilon (t) dt}$.

Note that while the holonomy is induced by slow changes of physical control parameters in 
adiabatic HQC, these parameters play a passive role in the non-adiabatic version. In particular, 
this means that the non-adiabatic scheme is not restricted to slow evolution and can therefore 
be made less exposed to decoherence effects by decreasing the run time of the gates 
\cite{johansson12}. 

\section{Spin chain model}
\label{sec:xymodel}
\begin{figure}[htp]
\centering
\includegraphics[width=10 cm]{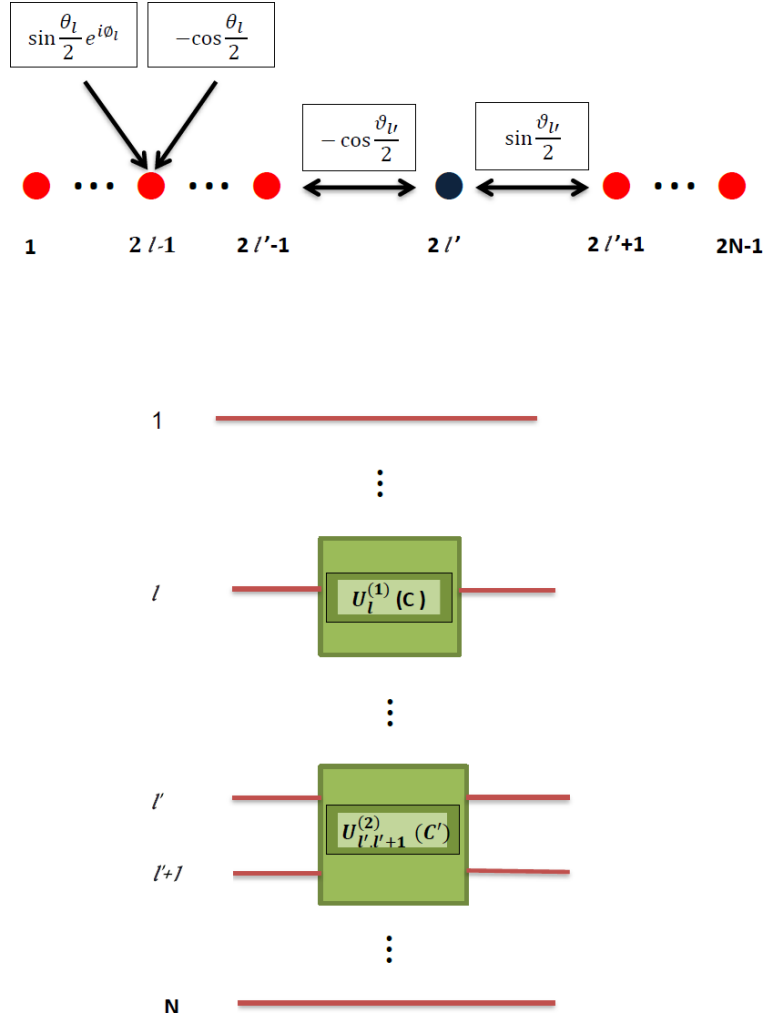}
\caption{Chain of three-level systems (upper panel) and its circuit equivalent (lower panel). 
The red-marked odd-numbered sites contain the logical qubits encoded in two-dimensional 
subspaces spanned by $\ket{0},\ket{1}$ of the internal three-level systems spanned by 
$\ket{0},\ket{1},\ket{e}$. In this way, $N$ qubits are obtained in a system of $2N-1$ three-level 
systems. A holonomic one-qubit gate $U_l^{(1)} (C)$ acting on qubit $l$ 
is realized by applying a $\pi$ pulse at site $2l-1$ of two coordinated laser fields that drive 
the $\ket{0} \leftrightarrow \ket{e}$ and $\ket{1} \leftrightarrow \ket{e}$ with relative 
phase $\phi_l$ and relative amplitude $-\tan (\theta_l/2)$. Similarly, a two-qubit 
gate $U_{l',l'+1}^{(2)}(C')$ acting on qubits $l'$ and $l'+1$ is realized by turning on interaction 
between pseudo-spins $2l'-1$, $2l'$, and $2l'+1$ only. The relative strength of the couplings 
is $-\tan (\vartheta_{l',l'+1})$. The holonomies $U_l^{(1)} (C)$ and $U_{l',l'+1}^{(2)} (C')$ 
are determined by the loops $C$ and $C'$ in the Grassmannian $G(3;2)$.}
\label{fig:fig1}
\end{figure}

Consider a linear chain of $2N-1$ three-level systems with controllable pair-wise nearest-neighbor 
isotropic XY-type interactions and local $\Lambda$ configurations driven by a pair of 
zero-detuned external fields (see upper panel of Fig.~\ref{fig:fig1}). The system evolution 
is governed by the Hamiltonian 
\begin{eqnarray}
H(t) & = & 
\sum_{k=1}^{N} f_k (t) \left[ \sin \frac{\theta_k}{2} \left( \cos \phi_k \lambda_{2k-1}^{(1)} - 
\sin \phi_k \lambda_{2k-1}^{(2)} \right) - \cos \frac{\theta_k}{2} \lambda_{2k-1}^{(4)} \right]
\nonumber \\ 
 & & + \frac{1}{2} \sum_{k=1}^{N-1} g_{k,k+1} (t) 
\left[ - \cos \frac{\vartheta_{k,k+1}}{2} \left( \lambda_{2k-1}^{(6)} \lambda_{2k}^{(6)} + 
\lambda_{2k-1}^{(7)} \lambda_{2k}^{(7)} \right) \right. 
\nonumber \\ 
 & & \left. + \sin \frac{\vartheta_{k,k+1}}{2}  
\left( \lambda_{2k}^{(6)} \lambda_{2k+1}^{(6)} + \lambda_{2k}^{(7)} \lambda_{2k+1}^{(7)} \right) 
\right] 
\nonumber \\  
 & = & \sum_{k=1}^{N} f_k (t) H_k^{(1)} + \sum_{k=1}^{N-1} g_{k,k+1} (t) H_{k,k+1}^{(3)} ,  
\end{eqnarray} 
where $H_k^{(1)},H_{k,k+1}^{(3)}$ are time independent during each pulse and the 
corresponding pulse and coupling envelopes $f_k (t),g_{k,k+1}(t)$ are real-valued. The 
relevant Gell-Mann operators associated with site $k$ read 
\begin{eqnarray}
\lambda_k^{(1)} & = & \ket{e}_k \bra{0}_k + \ket{0}_k \bra{e}_k , 
\nonumber \\ 
\lambda_k^{(2)} & = & -i\ket{e}_k\bra{0}_k + i\ket{0}_k \bra{e}_k , 
\nonumber \\ 
\lambda_k^{(4)} & = & \ket{e}_k \bra{1}_k +\ket{1}_k \bra{e}_k , 
\nonumber \\ 
\lambda_k^{(6)} & = & \ket{0}_k \bra{1}_k +\ket{1}_k \bra{0}_k , 
\nonumber \\ 
\lambda_k^{(7)} & = & -i\ket{0}_k \bra{1}_k + i\ket{1}_k \bra{0}_k , 
\end{eqnarray}
where $\ket{0},\ket{1},\ket{e}$ span the local state space. Each three-level system has 
a qubit subspace spanned by $\ket{0},\ket{1}$ with associated Pauli operators $\sigma_k^{x} = 
\ket{0}_k \bra{1}_k + \ket{1}_k \bra{0}_k$,  $\sigma_k^{y} = -i\ket{0}_k \bra{1}_k + i\ket{1}_k \bra{0}_k$, 
and $\sigma_k^{z} = \ket{0}_k \bra{0}_k - \ket{1}_k \bra{1}_k$ defining a pseudo-spin-$\frac{1}{2}$ 
system. Note that $\sigma_k^{x} = \lambda_k^{(6)}$ and $\sigma_k^{y} = \lambda_k^{(7)}$.  

A logical qubit is encoded in the two-dimensional subspace spanned by $\ket{0}$ 
and $\ket{1}$ of each odd-numbered three-level system. The auxiliary even-numbered systems 
act as a computational resource for mediating two-qubit gates, as will be shown below. 
In this way, $N$ logical qubits are obtained from the $2N-1$ systems (see lower panel of 
Fig.~\ref{fig:fig1}). The state space of the $N$ logical qubits $\mathcal{H}^{(N)}$ is spanned 
by the $2^N$ states 
\begin{equation}\{ \ket{n_1}_1 \ket{0}_2 \ket{n_2}_3 \ldots \ket{0}_{2N-2} \ket{n_N}_{2N-1}  
\equiv \ket{n_1 n_2 \ldots n_N}_L \}_{n_1,n_2,\ldots,n_N=0,1} \label{eqn5}
\end{equation} 
defined by setting the state of all auxiliary systems to $\ket{0}$.

The above Hamiltonian can be implemented in internal energy levels of atoms trapped 
in a one-dimensional optical lattice and exhibiting the desired XY-type interaction by adjusting 
the standing wave optical potential of the lattice \cite{duan03}. Another possible realization 
consists of ions trapped along a line by off-resonant standing waves. The internal states in 
this setting can be made to interact via state-dependent Stark shifts that couple the ions to 
the vibrational degrees of freedom of the trap \cite{porras04}. In both scenarios, pairs of 
zero-detuned laser fields couple each qubit level to an excited state, which is unaffected 
by the trapping fields. 

\section{Holonomic quantum computation}
\label{sec:hqc}
\subsection{One-qubit gates}
\label{sec:1qubit}
A holonomic one-qubit gate acting on qubit $l$ is implemented by turning on and off 
$f_l (t)$ at $t=0$ and $t=\tau$, respectively, and by simultaneously putting $g_{l-1,l} (t) = 
g_{l,l+1} (t) = 0$. By following Ref.~\cite{sjoqvist12}, we obtain the gate (see lower panel of 
Fig. \ref{fig:fig1})
\begin{eqnarray} 
U_l^{(1)} (C_{{\bf n}_l}) = {\bf n}_l \cdot \boldsymbol{\sigma}_l 
\end{eqnarray} 
acting on the $l$th qubit subspace spanned by $\ket{0},\ket{1}$, if the pulse area satisfies 
$\int_0^{\tau} f_l (t) dt = \pi$. Here, ${\bf n}_l = (\sin \theta_l \cos \phi_l\sin \theta_l 
\sin \phi_l, \cos \theta_l )$ and $C_{{\bf n}_l}$ is the loop in the Grassmannian $G(3;2)$ 
traversed by the qubit subspace. The geometric nature is guaranteed by noting that $H(t)$ 
vanishes on the qubit subspace on $[0,\tau]$. Two subsequent $\pi$ pulses with ${\bf n}_l$ 
and ${\bf m}_l$, yields 
\begin{eqnarray}
U_l^{(1)} (C_{{\bf m}_l}) U(C_{{\bf n}_l}) = {\bf n}_l \cdot {\bf m}_l - i\boldsymbol{\sigma} \cdot 
({\bf n}_l \times {\bf m}_l )
\end{eqnarray}
which proves the desired one-qubit universality. Thus, $U^{(1)}$ is just the holonomic 
one-qubit gate proposed in Ref.~\cite{sjoqvist12} applied to site $l$ of the chain.  

\subsection{Two-qubit gates}
\label{sec:3qubit}
We now demonstrate how a non-adiabatic holonomic two-qubit gate acting on logical qubits 
$l'$ and $l'+1$ can be implemented. This is achieved by turning on the coupling between 
systems $2l'-1,2l',$ and $2l'+1$ at $t=0$ and turning it off at $t=\tau$, while all other 
coupling terms remain zero. Thus, the $N$-body Hamiltonian reduces to 
\begin{eqnarray}
H(t) = g_{l',l'+1}(t) H_{l',l'+1}^{(3)}  
\end{eqnarray}
during this time interval. Note that the local excited states $\ket{e}_{2l'-1}$, $\ket{e}_{2l'}$, 
and $\ket{e}_{2l'+1}$ are fully decoupled by $H_{l',l'+1}^{(3)}$ from the subspace 
$\{ \ket{p}_{2l'-1} \ket{q}_{2l'} \ket{r}_{2l'+1} \equiv  \ket{pqr} \}$, $p,q,r=0,1$.   

The Hamiltonian $H_{l',l'+1}^{(3)}$ has four invariant subspaces 
$\mathcal{M}_{-\frac{3}{2}} = \{ \ket{111} \}$, $\mathcal{M}_{-\frac{1}{2}} = 
\{ \ket{011}, \ket{101}, \ket{110} \}$, $\mathcal{M}_{+\frac{1}{2}} = \{ \ket{100}, \ket{010}, 
\ket{001} \}$, and $\mathcal{M}_{+\frac{3}{2}} = \{ \ket{000} \}$, each labeled by the  quantum 
number $m=-\frac{3}{2},\ldots,+\frac{3}{2}$ of the total pseudo-spin operator $S_z =  
\frac{1}{2} \left( \sigma_{2l'-1}^z + \sigma_{2l'}^z + \sigma_{2l'+1}^z \right)$. Thus, 
\begin{eqnarray}
H_{l',l'+1}^{(3)} = \oplus_m H_{l',l'+1;m}^{(3)} , 
\end{eqnarray} 
where $H_{l',l'+1;m}^{(3)}$ acts on $\mathcal{M}_m$. This implies that the 
time evolution operator decomposes as 
\begin{eqnarray}
U (t,0) = \otimes_m U_m (t,0) = \otimes_m  e^{-i a_t H_{l',l'+1;m}^{(3)}} 
\end{eqnarray} 
with the `pulse area' $a_t = \int_0^t g_{l',l'+1}(t') dt'$. The computational states (defined 
by setting the auxiliary second qubit to $\ket{0}$) divide into $\mathcal{S}_1 = \{ \ket{000} 
\equiv \ket{00}_L \} = 
\mathcal{M}_{\frac{3}{2}}$, $\mathcal{S}_2 = \{ \ket{001} \equiv \ket{01}_L , \ket{100} 
\equiv \ket{10}_L \} \subset \mathcal{M}_{\frac{1}{2}}$, and $\mathcal{S}_3 = \{ \ket{101} 
\equiv \ket{11}_L \} \subset \mathcal{M}_{-\frac{1}{2}}$. Since $\mathcal{S}_1$ forms an 
invariant subspace, it undergoes a trivial evolution when exposed to $H_{l',l'+1}^{(3)}$. 
The Hamiltonian vanishes on $\mathcal{S}_2$ and $\mathcal{S}_3$, i.e., there will be 
no dynamical phases associated with the evolution of these two subspaces. On the other 
hand, $\mathcal{S}_2$ and $\mathcal{S}_3$ are proper subspaces of the invariant 
subspaces $\mathcal{M}_{\pm\frac{1}{2}}$ and may therefore pick up nontrivial 
holonomies. In this way, the evolution of $\mathcal{S}_1,\mathcal{S}_2,\mathcal{S}_3$ 
is purely geometric and defines holonomic two-qubit gates acting on the first and 
third site of the considered three-site block. We now demonstrate that these gates 
are in general sufficient to achieve universality when assisted by the above holonomic 
one-qubit gates.  

The parts $H_{l,l+1;\pm \frac{1}{2}}^{(3)}$ of the Hamiltonian, take the form
\begin{eqnarray}
H_{l,l+1;+\frac{1}{2}}^{(3)} & = & \sin \frac{\vartheta_{l',l'+1}}{2} \, \ket{010} \bra{001} - 
\cos \frac{\vartheta_{l',l'+1}}{2} \,  \ket{010} \bra{100} + \textrm{H.c.} , 
\nonumber \\ 
H_{l,l+1;-\frac{1}{2}}^{(3)} & = & \sin \frac{\vartheta_{l',l'+1}}{2} \,  \ket{101} \bra{110}  - 
\cos \frac{\vartheta_{l',l'+1}}{2} \,  \ket{101} \bra{011} + \textrm{H.c.} . 
\end{eqnarray}
Thus, $H_{l,l+1;+ \frac{1}{2}}^{(3)}$ exhibits a $\Lambda$-type configuration 
with $\ket{001}$ and $\ket{100}$ playing the role of the two `ground state' levels that couple  
to the `excited state' $\ket{010}$ with relative coupling strength $-\tan \frac{\vartheta_{l',l'+1}}{2}$.
The action of $U_{+\frac{1}{2}} (\tau,0)$ and $U_{-\frac{1}{2}} (\tau,0)$ on $\mathcal{S}_2$ 
and $\mathcal{S}_3$ using eqn. (\ref{eqn5}), respectively, is 
\begin{eqnarray}
A(a_{\tau}) & = & P_2 (0) U_{+\frac{1}{2}} (\tau,0) P_2 (0) 
\nonumber \\ 
 & = & \left( \cos^2 \frac{\vartheta_{l',l'+1}}{2}  + \sin^2 \frac{\vartheta_{l',l'+1}}{2} 
\cos a_{\tau}\right) \ket{01}_L \bra{01}_L 
\nonumber \\ 
 & & + \sin \vartheta_{l',l'+1} \sin^2 \frac{a_{\tau}}{2} \left( \ket{01}_L \bra{10}_L + 
\ket{10}_L \bra{01}_L \right) 
\nonumber \\ 
 & & + \left( \sin^2 \frac{\vartheta_{l',l'+1}}{2} + \cos^2 \frac{\vartheta_{l',l'+1}}{2} \cos a_{\tau} \right) 
\ket{10}_L \bra{10}_L , 
\nonumber \\ 
c(a_{\tau}) & = & P_3 (0) U_{-\frac{1}{2}} (\tau,0) P_3 (0) = \cos a_{\tau} \ket{11}_L \bra{11}_L ,
\end{eqnarray}
where $P_2(0)$ and $P_3(0)$ are projection operators onto $\mathcal{S}_2$ and $\mathcal{S}_3$, 
respectively. By chosing $a_{\tau} = \pi$, the evolution becomes cyclic and the resulting action 
of the time evolution operator on the computational 
subspace $\mathcal{S}_1 \oplus \mathcal{S}_2 \oplus \mathcal{S}_3$  becomes unitary and reads
\begin{eqnarray} 
U_{l',l'+1}^{(2)} (C') &  = & \ket{00}_L \bra{00}_L + A(\pi) +c(\pi)  
\nonumber \\ 
 & = & \ket{00}_L \bra{00}_L + \cos \vartheta_{l',l'+1} \ket{01}_L \bra{01}_L 
\nonumber \\ 
 & & + \sin \vartheta_{l',l'+1} \left( \ket{01}_L \bra{10}_L + \ket{10}_L \bra{01}_L \right) 
\nonumber \\ 
 & & - \cos \vartheta_{l',l'+1} \ket{10}_L \bra{10}_L - \ket{11}_L \bra{11}_L .
\end{eqnarray}
$U_{l',l'+1}^{(2)} (C')$ is the holonomic two-qubit gate acting on the qubit-pair $l',l'+1$ encoded 
in pseudo-spins $2l' \pm 1$ of the chain (see lower panel of Fig. \ref{fig:fig1}). $U_{l',l'+1}^{(2)} (C')$ is 
entangling and thereby universal when assisted by arbitrary one-qubit gates acting on the first 
and third qubit \cite{brenner02}. Thus, $U^{(1)} (C)$ and $U^{(2)} (C')$ constitute a universal, 
explicitly scalable set of holonomic gates associated with loops in $G(3;2)$. 

\section{Conclusions}
Quantum computation by holonomic means has attracted considerable attention in the past, 
due to the resilience of quantum holonomies to parameter fluctuations. In this paper, we have 
introduced a universal, explicitly scalable scheme based on non-Abelian holonomies (geometric 
phases) realized by interacting three-level systems, each of which exhibiting a $\Lambda$-type 
configuration. Our scheme provides an architecture for a compact holonomic quantum computer 
based on fast non-adiabatic evolution. This latter feature opens up for the possility to avoid 
detrimental open systems effects by decreasing the run time of the gates. The proposed setup 
may be implemented in atoms or ions trapped in tailored pulsed standing wave potentials. 

\section*{Acknowledgements}
This work was supported by the National Research Foundation and the Ministry of 
Education (Singapore). E.S. acknowledges support from the Swedish Research Council 
(VR) through Grant No. D0413201.


\begin{thebibliography}{99}
\bibitem{zanardi99} P. Zanardi, M. Rasetti,
Phys. Lett. A 264 (1999) 94.
\bibitem{pachos01} J. Pachos, P. Zanardi,
Int. J. Mod. Phys. B 15 (2001) 1257.
\bibitem{pachos00} J. Pachos, S. Chountasis,
Phys. Rev. A 62 (2000) 052318.
\bibitem{duan01} L. M. Duan, J. I. Cirac, P. Zoller,
Science 292 (2001) 1695.
\bibitem{pachos02} J. Pachos, H. Walther, 
Phys. Rev. Lett. 89 (2002) 187903. 
\bibitem{recati02} A. Recati, T. Calarco, P. Zanardi, J. I. Cirac, P. Zoller,
Phys. Rev. A 66 (2002) 0302309.
\bibitem{solinas03} P. Solinas, P. Zanardi, N. Zangh\`i, F. Rossi,
Phys. Rev. B 67 (2003) 121307.
\bibitem{bernevig05} B. A. Bernevig, S. C. Zhang,
Phys. Rev. B 71 (2005) 035303.
\bibitem{faoro03} L. Faoro, J. Siewert, R. Fazio,
Phys. Rev. Lett. 90 (2003) 028301.
\bibitem{brosco08} V. Brosco, R. Fazio, F. W. J. Hekking, A. Joye, 
Phys. Rev. Lett. 100 (2008) 027002. 
\bibitem{pirkkalainen10} J.-M. Pirkkalainen, P. Solinas, J. P. Pekola, M. M\"ott\"onen,
Phys. Rev. B 81 (2010) 174506.
\bibitem{wu05} L.-A. Wu, P. Zanardi, D. A. Lidar,
Phys. Rev. Lett. 95 (2005) 130501. 
\bibitem{renes13} J. M. Renes, A. Miyake, G. K. Brennen, S. D. Bartlett, 
New J. Phys. 15 (2013) 025020.
\bibitem{chancellor13} N. Chancellor, S. Haas, 
Phys. Rev. A 87 (2013) 042321. 
\bibitem{sjoqvist12} E. Sj\"oqvist, D. M. Tong, B. Hessmo, M. Johansson, K. Singh,
New J. Phys. 14 (2012) 103035.
\bibitem{anandan88} J. Anandan,
Phys. Lett. A 133 (1988) 171.
\bibitem{abdumalikov13} A. A. Abdumalikov, J. M. Fink, K. Juliusson, M. Pechal, S. Berger, 
A. Wallraff, S. Filipp, 
Nature 496 (2013) 482. 
\bibitem{feng13} G. Feng, G. Xu, G. Long, 
Phys. Rev. Lett. 110 (2013) 190501. 
\bibitem{arroyo-camejo14} S. Arroyo-Camejo, A. Lazariev, S. W. Hell, G. Balasubramanian, 
Nature Comm. 5 (2014) 4870.
\bibitem{zu14} C. Zu, W.-B. Wang, L. He, W.-G. Zhang, C.-Y. Dai, F. Wang, L.-M. Duan, 
Nature 514 (2014) 72.
\bibitem{xu12} G. F. Xu, J. Zhang, D. M. Tong, E. Sj\"oqvist, L. C. Kwek, 
Phys. Rev. Lett. 109 (2012) 170501.
\bibitem{zhang14a} J. Zhang, L. C. Kwek, E. Sj\"oqvist, D. M. Tong, P. Zanardi, 
Phys. Rev A 89 (2014) 042302.
\bibitem{liang14} Z.-T. Liang, Y.-X. Du, W. Huang, Z.-Y. Xue, H. Yan, 
Phys. Rev A 89 (2014) 062312.
\bibitem{xu14a} G. Xu and G. Long, 
Phys. Rev. A 90 (2014) 022323.  
\bibitem{xu14b} G. Xu and G. Long, 
Sci. Rep. 4 (2014) 6814.  
\bibitem{mousolou14} V. A. Mousolou, C. M. Canali, E. Sj\"oqvist, 
New J. Phys. 16 (2014) 013029.
\bibitem{zhang14b} J. Zhang, T. H. Kyaw, D. M. Tong, E. Sj\"oqvist, L. C. Kwek, 
arxiv:1412.2848.
\bibitem{bengtsson06} I. Bengtsson, K. \.{Z}yczkowski,
{\it Geometry of quantum states} (Cambridge University Press,
Cambridge, 2006) ch 4.9.
\bibitem{johansson12} M. Johansson, E. Sj\"oqvist, L. M. Andersson, M. Ericsson, B. Hessmo,
K. Singh, D. M. Tong,
Phys. Rev. A 86 (2012) 062322.
\bibitem{duan03} L.-M. Duan, E. Demler, M. D. Lukin, 
Phys. Rev. Lett. 91 (2003) 090402.
\bibitem{porras04} D. Porras, J. I. Cirac, 
Phys. Rev. Lett. 92 (2004) 207901.
\bibitem{brenner02} M. J. Bremner, C. M. Dawson, J. L. Dodd, A. Gilchrist, A. W. Harrow,
D. Mortimer, M. A. Nielsen, T. J. Osborne,
Phys. Rev. Lett. 89 (2002 247902.
\end{thebibliography}
\end{document}